# Large Language Models as AI Agents for Digital Atoms and Molecules: Catalyzing a New Era in Computational Biophysics

Yijie Xia[1,3], Xiaohan Lin[1,2,3], Zicheng Ma[4,5], Jinyuan Hu[1,3], Yanheng Li[1,3], Zhaoxin Xie[1,3], Hao Li[1,3], Li Yang[2], Zhiqiang Zhao[2], Lijiang Yang[1,3], Zhenyu Chen[1,2,3 a)], Yi Qin Gao[1,3,4 a)]

1. New Cornerstone Science Laboratory, College of Chemistry and Molecular Engineering, Peking University, Beijing 100871, China
2. Beijing Sidereus Intelligent Computing Technology Co., Ltd., Beijing 100080, China
3. Beijing National Laboratory for Molecular Sciences, Peking University, Beijing 100871, China
4. Changping Laboratory, Beijing 102206, China
5. Academy for Advanced Interdisciplinary Studies, Peking University, Beijing 100871, China

a) Authors to whom correspondence should be addressed: chemzyc@stu.pku.edu.cn and gaoyq@pku.edu.cn

ABSTRACT:

In computational biophysics, where molecular data is expanding rapidly and system complexity is increasing exponentially, large language models (LLMs) and agent-based systems are fundamentally reshaping the field. This perspective article examines the recent advances at the intersection of LLMs, intelligent agents, and scientific computation, with a focus on biophysical computation. Building on these advancements, we introduce ADAM (Agent for Digital Atoms and Molecules), an innovative multi-agent LLM-based framework. ADAM employs cutting-edge AI architectures to reshape scientific workflows through a modular design. It adopts a hybrid neural-symbolic architecture that combines LLM-driven semantic tools with deterministic symbolic computations. Moreover, its ADAM Tool Protocol (ATP) enables asynchronous, database-centric tool orchestration, fostering community-driven extensibility. Despite the significant progress made, ongoing challenges call for further efforts in establishing benchmarking standards, optimizing foundational models and agents, building an open collaborative ecosystem and developing personalized memory modules. ADAM is accessible at https://sidereus-ai.com.

# I. Introduction

In recent years, computational biophysics has undergone transformative development by advancement in algorithms, computing power, and data availability. There has been an explosion of computational methods and tools across all major subfields. Breakthroughs include predicting protein structures with atomic-level accuracy,[1-4] developing accelerated molecular docking frameworks capable of managing complex biological systems,[5-9] and improving molecular simulations through enhanced sampling techniques and the integration of AI technologies.[10-14]

The rapid expansion in computational biophysics has not only significantly advanced biophysical research frontiers but also introduced multidimensional complexities. A key barrier lies in the growing disconnect between specialized theoretical knowledge and the practical usage of advanced computational methods. This disconnect makes it difficult for many researchers to utilize cutting-edge tools effectively. This problem is worsened by the fragmented computational hardware and software ecosystems, which create steep learning curves and compatibility challenges for building efficient workflows, even for experienced researchers.. Moreover, the unstructured and multimodal nature of biophysical data, which often varies across experimental sources, analytical methods, and processing pipelines, requires intensive integration efforts and complicates systematic analysis.

At the same time, the development of large language models (LLMs) has marked a significant milestone in the field of artificial intelligence (AI). Initially applied to general language tasks like translation, information retrieval, and conversational agents, LLMs are now increasingly used in scientific domains.[15-18] They can assist with scientific question answering,[19-21] document classification,[22-24] and molecular property prediction.[20,25-27] Beyond traditional conventional natural language processing (NLP), LLMs can function as AI agents through tool calling and thus perform tasks autonomously.[28,29] In computational biophysics, these agents can gather and summarize biophysical knowledge from various sources, process diverse biophysical data, and utilize different computational tools. They also aid experimental scientists in handling unstructured biophysical data and executing complex physical simulations.

This perspective article first summarizes recent advancements in LLMs and agent-based systems, with a focused analysis of their applications in computational biophysics. We also present the Agent for Digital Atoms and Molecules (ADAM), a transformative framework for computational biophysics. We provide a systematic introduction to ADAM's overall architecture, highlighting its hybrid neural-symbolic tool systems and community-driven extensibility protocol. Finally, we discuss key challenges and future directions for agent development in this field.

# II. Computational Biophysics in the LLM Era

## A. Recent advancements in LLMs

LLMs are large-scale, pre-trained statistical language models based on neural network architectures.

Their emergence marks a major milestone in AI research and applications.[30-33] Well-known LLMs like GPT,[34,35] LLaMA,[36] and DeepSeek[37,38] are making big impacts across many fields. Most of these models use the transformer architecture,[39] which is key in modern NLP. This architecture can capture long-range dependencies and rich context. Pretraining on extensive web-scale text data allows LLMs to internalize complex linguistic patterns and world knowledge. As a result, LLMs perform exceptionally well in many NLP applications such as machine translation, speech recognition, information retrieval, question answering, customer support, and conversational systems.[40-42]

Recent innovations like Multi-head Latent Attention,[38] Mixture-of-Experts architectures,[43] and Multi-Token Prediction mechanisms[44] have effectively solved critical computational efficiency, long-context processing, and scalability issues in many scenarios. LLMs' reasoning capabilities have also been greatly improved by advanced inference techniques such as Chain-of-Thought (CoT)[45] prompting and Monte Carlo Tree Search[46]. As inference costs drop and performance improves, LLMs are becoming more accessible for consumer-level use and real-world practical applications. Meanwhile, LLMs have become essential scientific tools, reshaping workflows across disciplines and changing traditional scientific inquiry paradigms.

## B. Direct applications of LLMs in computational biophysics

The unique capabilities of LLMs in understanding and generating complex sequences have created exciting opportunities in computational biophysics. This trend is particularly evident in their application to the long-standing challenges in protein structure prediction. Traditional approaches to this problem include molecular dynamics (MD), Monte Carlo simulations, and Markov random fields. [47-53] These methods rely on physics-driven energy landscape sampling to infer protein conformations, but are computationally expensive and struggle with exploring high-dimensional conformational spaces. Thanks to the exponential rise in experimentally resolved protein structures and deep learning innovations, this field has been greatly changed. Frameworks based on language models, such as AlphaFold2[2] and ESMFold,[54] now achieve prediction accuracies close to experimental precision.

Besides the success in macromolecular modeling, LLMs are also extending their impact to small-molecule systems. In small molecule property prediction, traditional computational methods often involve explicit calculations of atomic interactions through MD or density functional theory (DFT).[55,56] However, advanced LLMs have shown significant potential in predicting molecular properties and in designing novel molecular structures. For example, SELFormer[27] uses SELFIES language models to predict small molecule properties. This extension of LLMs from macromolecular to small-molecule systems highlights their growing influence across different scales of molecular research.

## C. Augmenting LLMs through external knowledge

Though LLMs show broad and promising capabilities, critical limitations hinder their reliability in scientific applications. LLMs are predominantly trained on static, fixed-timepoint datasets, rendering them prone to knowledge obsolescence - a significant issue in fast-evolving fields like computational biophysics where new data and methods emerge frequently. Also, they are prone to generating hallucinations, producing content that's nonsensical or inconsistent with sources, thus leading to factual errors.[57] Moreover, their performance in specialized fields is limited due to insufficient domain-specific training, affecting their effectiveness in dynamic, complex systems. To address these limitations, researchers are pursuing two complementary strategies: specializing LLMs through targeted post-training, and augmenting them with real-time external knowledge, which offer distinct pathways to enhance scientific reliability.

The first strategy focuses on domain specialization through post-training methodologies. The related methods include supervised fine-tuning techniques,[29,58] reinforcement fine-tuning[59,60] approaches and test-time scaling[61-63] methods. A recent study showed that fine-tuning Qwen2.5-32B-Instruct with just 817 carefully chosen examples allowed it to outperform existing state-of-the-art models, including ChatGPT-o1, in solving complex mathematical competition problems.[64] This research highlights that combining high-quality, domain-specific training data with well-structured prompts during inference can effectively activate and enhance the domain-relevant reasoning abilities within LLMs. The targeted activation improves autonomous reasoning and significantly boosts model performance in specialized tasks.

The second strategy employs retrieval-augmented generation (RAG), which dynamically integrates external knowledge to overcome the limitations of static training data.[65] RAG enhances LLM outputs by retrieving relevant external information as context, thereby improving accuracy and richness of generated content. Beyond mitigating outdated knowledge, RAG is frequently integrated with external resources for efficient domain-specific task handling, and it offers a decentralized, privacy-preserving framework ideal for individual knowledge base systems and privacy-aware agents.

The evolution of RAG spans multiple stages.[66] Initially, Early implementations relied on basic keyword-based retrieval techniques, such as TF-IDF and BM25,[67,68] but these naïve RAG systems had limited contextual awareness, fragmented outputs, and poor scalability. Advanced RAG systems address these shortcomings through sophisticated text vectorization techniques and hybrid retrieval strategies.[69] In addition, re-ranking[70] and multi-hop retrieval[71] mechanisms are incorporated to refine contextual precision and accuracy. Innovative variants further expand RAG's capabilities. Graph-based RAG[72] utilizes hierarchical graph structures to manage structured and unstructured data, enhancing entity relationship modeling and knowledge graph traversal. Modular RAG[73] decouples the retrieval-generation pipeline into independent and reusable modules, allowing for domain-specific optimization and enhanced task adaptability.

## D. Agents

An agent is formally defined as an autonomous computational entity equipped with sensor-driven environmental perception, state interpretation, and action selection capabilities.[74-76] This operational autonomy enables self-directed decision-making and iterative workflow optimization. Multi-agent systems consist of multiple LLM-based agents and modules, characterized by complex, dynamic interactions similar to human teamwork. They typically include an agent-environment interface, agent profiling for role-specific tasks, and communication mechanisms for information exchange.[76] In these systems, the planning agent breaks down complex queries into parallelizable subtask workflows. It uses critic modules and reflection mechanisms to iteratively assess intermediate results and dynamically adjust execution strategies. Meanwhile, the routing agent manages task delegation across specialized downstream agents.[77] Additionally, memory modules capture and maintain contextual information, knowledge, and agent experiences throughout interactions.[66] Multi-agent systems often integrate advanced tool-calling capabilities like vector search, web queries, custom functions, and API integrations. The combination of RAG and multi-agent frameworks shows great potential in applications such as software development,[78] social simulations[79] and gaming.[80]

The use of intelligent agents in scientific computing workflows is a major step towards automating research tasks, which can greatly boost research efficiency and productivity by handling time-consuming, cross-discipline, and labor-intensive tasks. Recent studies have delved into the abilities of LLM-based intelligent agents in automating computational biophysics workflows. For example, MDCrow[81] is an intelligent agent that excels at managing complex MD simulations, enabling researchers to tackle intricate and computationally intensive systems. Additionally, BioAgents[82], a multi-agent system that combines reinforcement fine-tuning and RAG, has achieved expert-level performance on conceptual genomic tasks.

The integration of intelligent agents into database-related contexts has catalyzed the emergence of next-generation visual analytics platforms, transforming how researchers interact with complex biophysical datasets. For example, DrBioRight[83] uses intelligent agents to automate procedures and visualization in a cancer proteomics database, enhancing usability and accessibility for researchers.

These scientific computing agents extend beyond workflow automation to serve as cognitive collaborators. This capability fundamentally redefines human-computer interaction in computational biophysics. Previously, extensive documentation and specialized knowledge were required to operate complex software as well as hardware. Now, natural language interactions make these tools more accessible, reducing entry barriers and providing researchers including experimentalist and theoreticians with easily accessible support.

# III. ADAM: A Multi-Agent Framework for Collaborative Biophysical Computation

## A. Overall Framework

The Agent for Digital Atoms and Molecules (ADAM) introduced here represents a multi-agent framework that combines LLMs with existing scientific tools to address complexity and fragmentation in computational biophysics. ADAM's adaptive architecture classifies computational biophysical queries into general and specific tiers based on their operational complexity and technical requirements.

General inquiries tackle open-ended research challenges requiring cross-domain integration. A typical scenario is given as the following example: "Design an antibody with optimized CDRs to maximize binding affinity for the uploaded target protein (PDB: 8ABC)." These requests require coordinated cross-domain and multi-stage computational workflows - starting with AI-driven structural prediction, moving to physics-based molecular docking, and ending with molecular dynamics for free energy landscape sampling to comprehensive validation of binding interactions.

Some specific inquiries involve well-defined technical procedures, exemplified by: "Execute rigid-body docking between the submitted protein (PDB: 1XYZ) and ligand (SMILES: C1=CC(=CC=C1F)Cl) using DSDP." Such operations focus on executing standardized computational workflows or confirming experimental results using proven algorithmic tools.

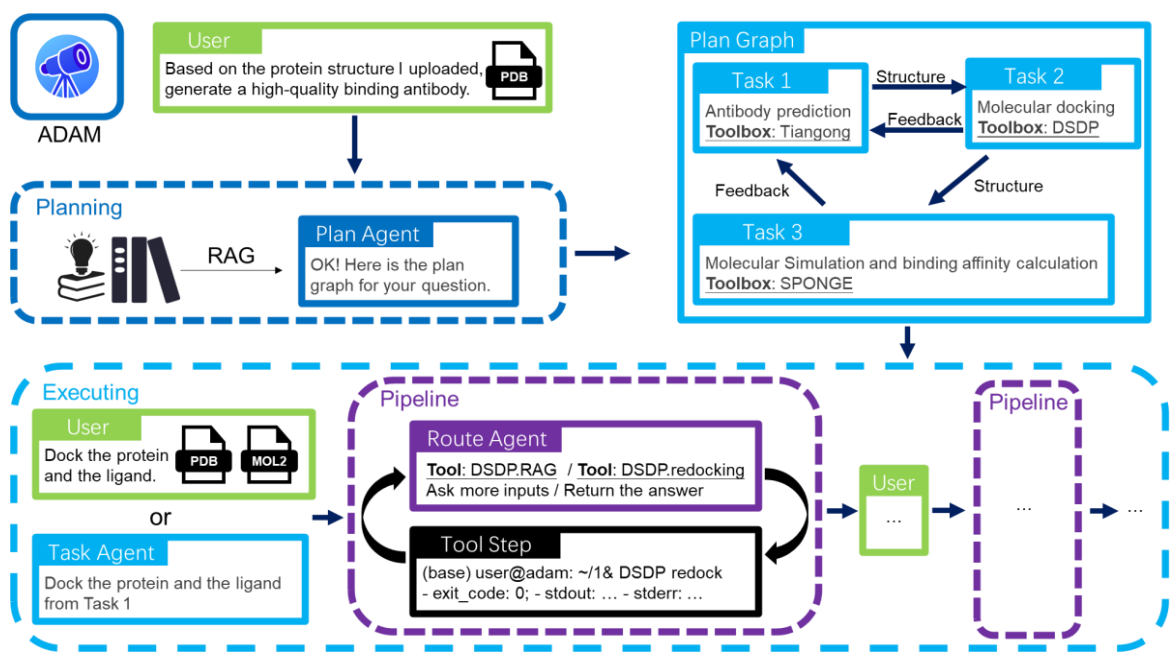

**Fig 1**. The overview of the framework for ADAM.

As illustrated in Fig. 1, ADAM's architecture is designed to operate through a coordinated multi-agent interaction framework. For general problem-solving, the plan agent functions as the cognitive core. It parses user input and constructs hierarchical plan graphs through question decomposition. This process translates complex biophysical questions into sequential technical operations matched to ADAM's toolboxes.

For either inquiries decomposed from the plan agent or direct user-defined atomic directives, the route agent dynamically orchestrates execution pipelines. During execution, it "intelligently" selects domain-specific tools or generates real-time responses. Post-execution result analysis passes to the route agent, which then decides to either iteratively deploy subsequent tools or terminate the pipeline with consolidated outputs. Short-term memory is maintained through session-specific interaction logs, which store detailed records of user-system dialogues to enable low-latency retrieval of contexts. For long-term memory, the route agent summarizes historical interaction contexts to mitigate context window overflow.

## B. Hybrid Neural-Symbolic Architecture

In cognitive psychology, dual-process theory[84] delineates two distinct modes of information processing that govern human decision-making: System 1 (Intuitive Processing) and System 2 (Analytical Processing). From an evolutionary perspective, System 1 emerged as an energy-efficient mechanism for rapid survival-critical responses, while System 2 developed later to handle novel complex problems through conscious analysis. Modern neuroimaging studies[85-87] link System 1 operations to the basal ganglia and amygdala, and System 2 to prefrontal cortical networks.

In ADAM, this dichotomy is reflected in the neural and symbolic tools within its toolboxes, as shown in Fig. 2. ADAM's hybrid neural-symbolic architecture integrates neural networks for semantic understanding with symbolic systems for deterministic computations. This framework coordinates complementary tools within each task toolbox, balancing the flexibility of neural approaches with the scientific rigor of symbolic systems. Through this hybrid neural-symbolic integration, ADAM synergistically balances exploratory adaptability for open-ended biophysical discovery with deterministic precision in structured biophysical operations, ensuring both scientific creativity and computational reproducibility.

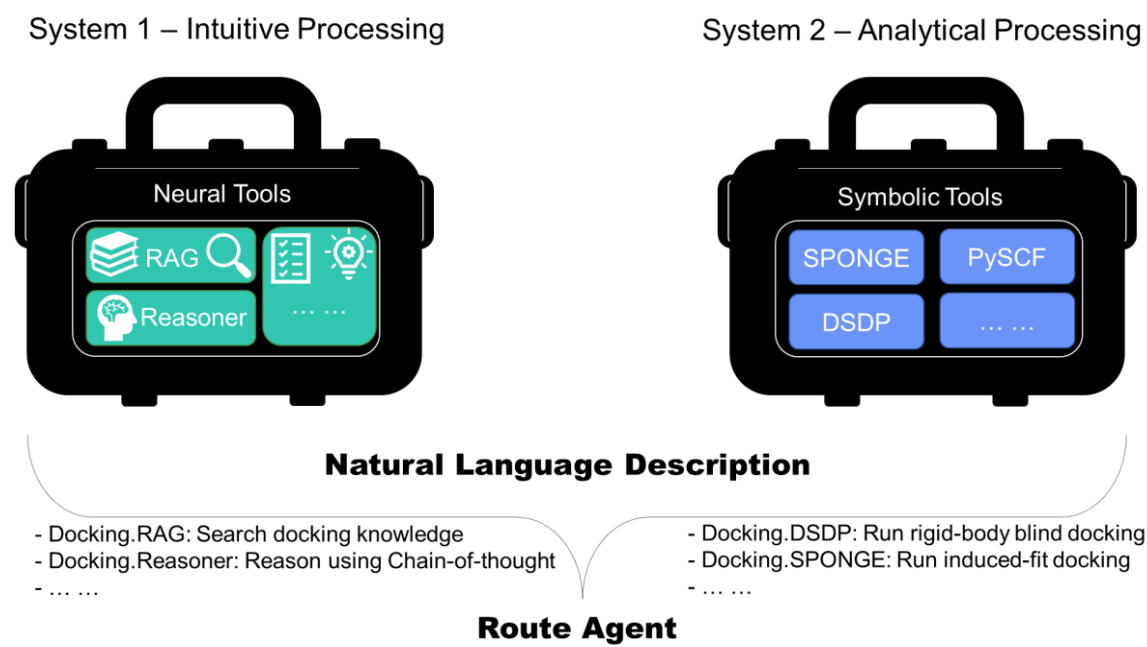

**Fig 2**. Illustration of the hybrid neural-symbolic architecture for the ADAM tools

The symbolic components in ADAM implement rigorously validated scientific computations via dedicated programmatic modules, executing domain-specific operations with deterministic precision. For example, molecular docking uses the DSDP[7] package for physics-based rigid-body alignment simulations, MD employs SPONGE[13] for trajectory calculations, and electronic structure analysis utilizes DFT in PySCF[88] for quantum mechanical property computations.

The neural components leverage specialized LLMs tailored for distinct unstructured tasks. For instance, the RAG Engine employs domain-specific knowledge retrievals with dynamic algorithm selection based on knowledge structures. Another example is the Chain-of-Thought Reasoner, which employs task-optimized fine-tuned LLMs for logically constrained multi-step inference. ADAM's semantic-driven tools enforce task-domain confinement—restricting LLMs and retrieval systems to predefined biophysical subdomains.

The route agent serves as the function dispatcher that bridges natural language interfaces with the computational tools through a standardized tool protocol. This protocol abstracts both neural components and symbolic tools as API-callable functions with self-documented natural language description of their capabilities. This bi-directional translation framework enables seamless integration of neural-symbolic operations and real-time assembly of hybrid toolchains through intent-aware routing.

## C. ADAM Tool Protocol

The ADAM architecture achieves systematic extensibility through its ADAM Tool Protocol (ATP), a standardized interoperability framework designed for seamless integration of third-party biophysical computation tools. As depicted in Fig. 3, ATP employs a server-executor architecture comprising three core components: an ATP service endpoint (server), distributed tool executors (clients), and a PostgreSQL-mediated communication layer.

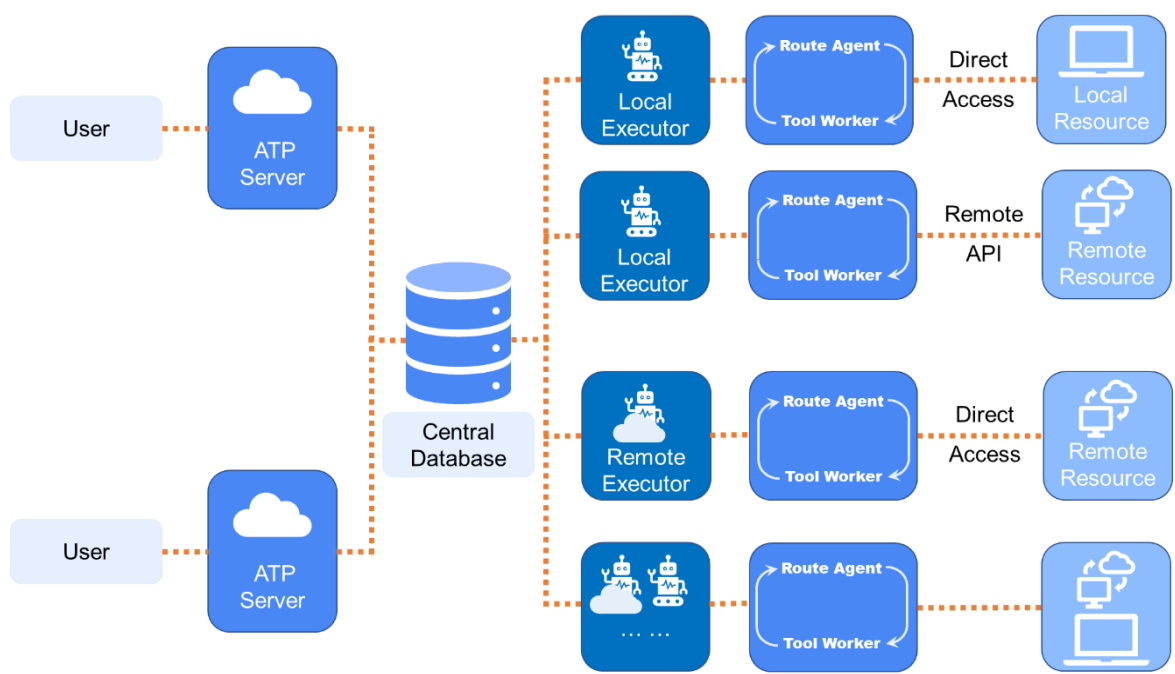

**Fig 3**. Illustration of the architecture for ATP

While general-purpose protocols like Model Context Protocol (MCP)[89,90] proposed by Anthropic and Agent-to-Agent (A2A) Protocol[91] developed by Google share similar objectives in enabling basic tool integration, ATP specifically addresses unmet needs in computational biophysics, including challenges unique to resource-constrained environments and domain-specific workflow demands.

ATP employs a database-centric communication architecture, leveraging the PostgreSQL wire protocol[92] through a shared database infrastructure. This design eliminates dependencies on dedicated public IP addresses for tool executors, directly resolving IPv4 address scarcity challenges, which are particularly prominent in resource-constrained networks common to developing countries like China.[93]

The protocol naturally enables asynchronous task management, decoupling job submission from execution to enhance scalability. ADAM's hybrid resource orchestration leverages local executors to directly interface with on-premise high-performance computing infrastructure while utilizing remote executors to smoothly integrate cloud-based distributed systems, enabling adaptive workload distribution across heterogeneous computational environments. A practical implementation of this architecture is demonstrated through our phased deployment. The initial integration phase established ADAM's local executor on a Dell Precision 7960 workstation (2×RTX4090 GPUs). With scale-out demands, the system transparently extends computation to PARATERA Supercomputing Cloud through remote executors. Notably, local executors retain the ability to access remote resources via standardized APIs, enabling cross-platform interoperability.

With designs, ATP is aimed to provide a robust, domain-optimized solution for third-party tool integration. Its server-executor architecture, database-driven communication, asynchronous task handling, and hybrid resource management collectively address limitations of general-purpose standards. By aligning with the technical and infrastructural realities of computational biophysics, ATP lowers barriers to community-driven tool adoption, fostering collaborative ecosystem growth for the ADAM platform.

# IV. Challenges and Future Directions

## A. Benchmarking standards establishment

Despite recent advancements in AI agents for computational biophysics (e.g., ADAM), critical limitations persist in handling multi-component systems, autonomous error correction, and cross-domain knowledge transfer. To systematically evaluate agent capabilities, we propose a six-tier competency framework (Table 1) that maps agent proficiency to academic maturity stages - from basic task execution (L0: Bachelor) to autonomous cross-domain framework innovation (L6: Genius).

**Table 1. AI Agent Capability Levels and Core Competency Framework**

| Level | Expertise Level | Competency Milestones |
|---|---|---|
| L0 | Undergraduate | Execute tasks with fully configured inputs |
| L1 | Graduate Student/Master | Perform post-processing on structured outputs |
| L2 | PhD Student/Junior | Generate input files and debug basic error |
| L3 | PhD Student/senior | Resolve complex conflicts through parameter tuning |
| L4 | PhD Graduate | Select optimal algorithms for complex tasks |
| L5 | Professor | Formulate hypotheses and design novel solutions |
| L6 | Genius | Create cross-domain frameworks with self-evolving capability |

Taking MD simulations as an example, while current agents can execute preconfigured molecular dynamics (MD) simulations (L0) and perform predefined trajectory analyses (L1), autonomous generation of chemically valid topology files remains confined to predefined systems at L2. Advancing further poses significant barriers: real-time sampling optimization (L3) struggles with multi-component phase equilibria and achieving L4 adaptive sampling necessitates simultaneous coordination of enhanced sampling algorithms and data-driven collective variable discovery for the agents. Current agents predominantly operate at L1–L2 proficiency, with none reaching L6's cross-domain capabilities. Similar limitations persist in other computational biophysics subfields, where agents rarely exceed L3 proficiency, underscoring the necessity for foundational algorithmic advancements to bridge these capability gaps.

**Table 2. Domain-Specific Proficiency in Molecular Simulation Mapped to AI Agent Capability Levels**

| Level | Exemplary Molecular Simulation Capabilities |
| --- | --- |
| L0 | Execute MD simulations using prepared topology/conformation files |
| L1 | Assess energy landscapes via automated trajectory analysis |
| L2 | Generate system-specific topology/conformation files with error correction |
| L3 | Optimize sampling strategies through real-time feedback |
| L4 | Implement adaptive enhanced sampling in multi-component systems |
| L5 | Develop transferable force fields via active learning |
| L6 | Integrate molecular dynamics into cross-domain frameworks (e.g., autonomous coupling with drug discovery or systems biology pipelines), enabling self-evolving hypothesis generation across disciplines. |

To drive progress, level-specific benchmarking standards must be established, defining quantitative thresholds and validation protocols for each capability tier. Such benchmarks would not only quantify agent capabilities but also guide targeted improvements in underperforming tiers, ensuring systematic progression toward autonomous, cross-domain computational frameworks.

## B. Foundational models and agents optimization

To achieve higher-tier agent capabilities, both the underlying LLMs and agent architectures require further systematic optimization. A persistent challenge lies in hallucination mitigation. LLMs occasionally generate factually incorrect outputs that evade detection in complex applications. While ADAM's hybrid neural-symbolic architecture and task decomposition framework reduce hallucinations through modular knowledge containment, hallucinations still persist, and further efforts are necessary. Some of the possible solutions are discussed below.

To advance the planning agent's capabilities as the cognitive core of ADAM, future work will prioritize synergistic research directions aimed at enhancing workflow orchestration and decision-making reliability. First, we propose to develop graph-based task representations informed by human-curated biophysical ontologies, where hierarchical knowledge graphs will encode probabilistic relationships between tasks to enable context-aware reasoning over multi-step workflows. Second, human-in-the-loop reinforcement learning frameworks will be implemented to allow domain experts to iteratively refine agent behavior through real-time feedback on intermediate outputs.

Evolving neural-symbolic systems into causal-neural-symbolic[94,95] architectures has the potential to significantly improve error traceability through explicit causal modeling. By embedding causal graphs, agents can perform root-cause analysis on aberrant outputs. This causal linkage not only improves the interpretability of results but also enables the agent system to self-adjust for more optimal outcomes.

In addition to the aforementioned improvements, integrating domain-specific tokenization schemes holds the potential to enable agents to better interpret multi-modal outputs from computational tools in tasks like binding pocket prediction or reaction pathway visualization. For example, current agents generally analyze protein structures merely by reading coordinate data from files, lacking the ability to truly grasp the three-dimensional architecture. In contrast, by employing ProTokens[96] - a domain-specific tokenization framework - these agents can achieve a deeper understanding of protein structures at both the symbolic and spatial levels. It is the difference between viewing a protein as a static list of coordinates and seeing it as a dynamic, three-dimensional entity with regions of interest and functional sites.

## C. Open collaborative ecosystem construction

Addressing the persistent fragmentation in computational biophysics demands the establishment of an open collaborative ecosystem that synergizes data, tools, and expertise across disciplines.

Historically, the field has been hindered by incompatible toolchains - many computational utilities generate non-standardized outputs and lack machine-readable documentation, while invaluable expert-novice knowledge exchanges remain uncaptured. To mitigate these challenges, future efforts should prioritize authenticated knowledge-sharing platforms that dynamically archive multi-modal interactions - including annotated troubleshooting logs, simulation trajectory annotations, and consensus-driven protocol optimizations - into structured, AI-ready corpora.

We hope that AI-adaptive toolchain standardization can represent a strategic advancement for fostering interoperability within computational biophysics ecosystems. Building on ATP, such standardization aims to unify domain-specific tools through cloud-native containerization, which may significantly reduce environment configuration complexities and enhance reproducibility across distributed research teams. By prioritizing modular and extensible design principles, this approach seeks to mitigate the inefficiencies of isolated workflows, enabling more cohesive integration of heterogeneous computational methods. If successfully implemented, these efforts could gradually elevate AI agents from specialized utilities to integral components of multidisciplinary research infrastructures, thereby supporting broader methodological convergence in computational biophysics.

## D. Personalized Agent Development with Individualized Memory

One of the most promising future directions in agent development involves creating personalized agents equipped with individualized long-term memory modules. While current agent systems can capture and summarize interaction contexts through basic memory functions, they lack the capacity to deeply integrate a user's unique knowledge base, experiential history, and cognitive frameworks. Implementing persistent memory architectures would enable agents to develop user-specific adaptations, thereby supporting lifelong learning trajectories and professional development - particularly crucial for educational applications in computational-biophysics training.

The development of personalized agents with individualized memory capabilities has been actively explored across multiple domains. For instance, in the field of healthcare, the "ReMe" framework[97] demonstrates that the personalized cognitive training with individualized memory modules can potentially mitigate cognitive decline in early-stage Alzheimer's patients. The gaming industry has similarly embraced this concept, incorporating player-specific memory systems into non-playable characters to enhance gameplay experience. [98,99] Furthermore, AI-powered personalization algorithms have seen widespread implementation, particularly in social and entertainment platform (e.g., TikTok's recommendation algorithm[100]), showcasing the practical viability of memory-based personalization at scale.

While these advancements across diverse fields highlight the promise of personalized agents equipped with memory modules, substantial challenges remain in adapting such systems for computational biophysics agents. From a technical perspective, developing these agents requires advancing sophisticated context-aware memory systems capable of processing longitudinal interaction data—such as students' problem-solving trajectories in molecular modeling. To train memory encoding mechanisms effectively, large-scale, domain-specific training datasets rooted in computational biophysics concepts must be curated. Additionally, a robust architectural framework is required to integrate user-specific knowledge backgrounds. Overall, building an agent that can dynamically tailor outputs to individual comprehension levels is a complex task requiring further development.

## V. Conclusion

The integration of LLMs and AI-driven multi-agent systems represents a potentially transformative paradigm for computational biophysics, offering solutions to persistent challenges of workflow fragmentation, data complexity, and accessibility barriers. The ADAM framework exemplifies this transformation through its hybrid neural-symbolic architecture and its community-driven extensibility protocols that enable seamless third-party tool integration. By deploying modular agents for context-aware task decomposition, dynamic tool orchestration, and persistent memory management, ADAM has the potential to reduce the gap between research planning and execution, as well as between theoretical modeling and experimental validation, empowering researchers to navigate complex biophysical inquiries through natural language interfaces without sacrificing methodological precision.

Current limitations highlight critical methodological gaps requiring focused attention: the absence of level-specific benchmarking standards, underdeveloped optimization strategies for LLM-agent architectures, and fragmented ecosystem interoperability. Future effort should prioritize addressing these challenges through systematic validation, algorithm refinement, community-driven standardization and user-specific memory optimization. By centering efforts on these objectives, frameworks like ADAM could mature into reliable tools for innovative research, strategically augmenting human expertise to navigate the multi-scale complexity inherent to computational biophysics.

# ACKNOWLEDGMENTS

This work was supported by the National Science and Technology Major Project (2022ZD0115003), the National Natural Science Foundation of China (No. 92353304, No. T2495221), and New Cornerstone Science Foundation (NCI202305). The AI-driven experiments, simulations and model training were performed on the robotic AI-Scientist platform of Chinese Academy of Sciences.

# AUTHOR DECLARATIONS

## Conflict of Interest

Xiaohan Lin and Zhenyu Chen are founders of Beijing Sidereus Intelligent Computing Technology Co., Ltd.

## Author Contributions

**Yijie Xia**: Conceptualization; Methodology; Investigation; Writing – original draft; Writing – review & editing; Software. **Xiaohan Lin**: Conceptualization; Methodology; Investigation; Writing – review & editing; Software. **Zicheng Ma**: Writing – original draft; Writing – review & editing; Software; **Jinyuan Hu**: Software. **Yanheng Li**: Software. **Zhaoxin Xie**: Software. **Hao Li**: Software. **Li Yang**: Software. **Zhiqiang Zhao**: Software; **Lijiang Yang**: Supervision, Project

Administration, Resources, Funding acquisition. **Zhenyu Chen**: Conceptualization; Methodology; Investigation; Writing – review & editing; Project Administration, Resources. **Yi Qin Gao**: Conceptualization; Supervision, Writing – review & editing, Project Administration, Resources, Funding acquisition.

# DATA AVAILABILITY

Data sharing is not applicable to this article as no new data were created or analyzed in this study.